\documentclass[twocolumn,english,superscriptaddress]{revtex4-1}
\usepackage[T1]{fontenc}
\usepackage[utf8]{inputenc}
\usepackage{color}
\usepackage{tipa}
\usepackage{tipx}
\usepackage{graphicx}
\usepackage{amsmath}
\usepackage{soul}
\makeatletter
\usepackage{babel}
\usepackage[ddmmyyyy]{datetime}
\makeatother

\let\a=\alpha \let\b=\beta \let\g=\gamma 
\let\e=\varepsilon \let\z=\zeta  \let\k=\kappa
\let\l=\lambda \let\m=\mu   
\let\s=\sigma   
   
\let\D=\Delta   
   
 \let\r=\rho \let\th=\theta

\def\MM{{\cal M}}

\def\DD{{\cal D}}

\def\to{\rightarrow}

\def\de{\mathrm d}

\newcommand{\beq}{\begin{equation}} 
\newcommand{\eeq}{\end{equation}}
\newcommand{\ba}{\begin{eqnarray}}
\newcommand{\ea}{\end{eqnarray}}

\def\de{\mathrm d}

\newcommand{\MeasureWP}[1]{ \DD' \underline w  }

\usepackage{babel}
\begin{document}
\title{Critical jammed phase of the linear perceptron}
\author{Silvio Franz}
\affiliation{LPTMS, Université Paris-Sud 11, UMR 8626 CNRS, Bât. 100, 91405 Orsay Cedex, France}
\affiliation{Dipartimento di Fisica Università, La Sapienza, Piazzale Aldo Moro 5, I-00185 Roma, Italy}
\author{Antonio Sclocchi}
\affiliation{LPTMS, Université Paris-Sud 11, UMR 8626 CNRS, Bât. 100, 91405 Orsay Cedex, France}
\author{Pierfrancesco Urbani}
\affiliation{Institut de Physique Théorique, Université Paris
Saclay, CNRS, CEA, F-91191, Gif-sur-Yvette, France}

\begin{abstract}
  Criticality in statistical physics naturally emerges at isolated
  points in the phase diagram. Jamming of spheres is not an exception:
 varying density, it is the critical point that separates the unjammed phase where
  spheres do not overlap and the jammed phase where they cannot be
  arranged without overlaps. The same remains
  true in more general constraint satisfaction problems with
  continuous variables (CCSP) where jamming coincides with the
  (protocol dependent) satisfiability transition point.  
In this work we show that by carefully choosing the cost function to be minimized, the region of criticality extends to occupy a whole region of the jammed
phase.  As a working example, we consider the
  spherical perceptron with a linear cost function in the unsatisfiable (UNSAT)
  jammed phase and we perform numerical simulations which show
  critical power laws emerging in the configurations obtained
  minimizing the linear cost function. We develop a scaling theory to
  compute the emerging critical exponents.
 \end{abstract}

\maketitle
{\it Introduction --}
The jamming transition of spheres is a critical point \cite{LN10}.
At jamming, spheres form an \emph{isostatic} network \cite{TW99} where the number of 
contacts between them equals exactly the total number of degrees of freedom.
Furthermore, the distributions of forces and gaps (i.e. distances) between particles display power laws \cite{Wy12, LDW13, MW15, CCPZ15}
which play a central role in the mechanical and rheological properties of such systems.
In \cite{CKPUZ14NatComm, CKPUZ14JSTAT} the corresponding critical exponents have been computed from the solution
of the hard sphere model in infinite dimension. 
In this analysis the jamming transition is thought of as the infinite pressure limit of hard sphere glassy states. 
\begin{figure}[h]
\centering
\includegraphics[width=\columnwidth]{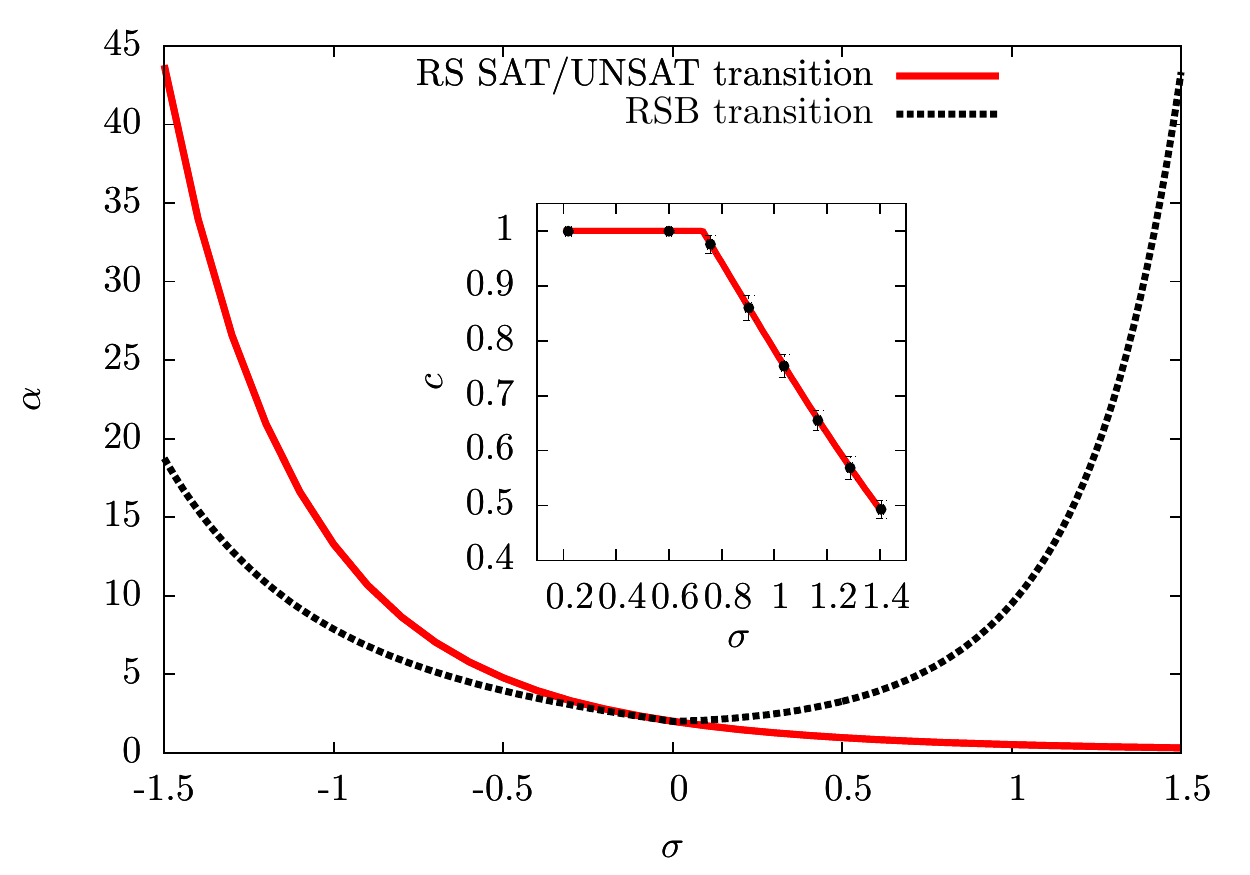}
\caption{The phase diagram of the spherical perceptron with linear cost function. 
The $y$-axis represents the density of constraints $\alpha=M/N$ while the $x$-axis is the control parameter $\sigma$ that defines the gap variables, see Eq.~\eqref{eq_gap}.
The red solid line is the SAT/UNSAT (jamming) line (computed under
the replica symmetric approximation, exact only for $\sigma>0$). 
Below this line the model is SAT/unjammed and one can find
zero energy configurations. Above, the problem is UNSAT/jammed and the energy is positive.
Above the black dashed line replica symmetry spontaneously breaks down:
below this line the problem is convex while
above the energy landscape is glassy with many local minima
with critical properties. The jamming line lies in the RS
region for $\sigma>0$ and in the RSB region for $\sigma<0$.
\emph{Inset}: The density of contacts $c$
for $\alpha=5$ as a function of $\sigma$. The red line corresponds to
the theoretical prediction. We have hypostaticity $c<1$ in the RS phase and isostaticity, i.e. $c=1$ in the RSB phase. The dots come from numerical simulations with $N=1024$ at $\a=5$.}
\label{Fig:PD}
\end{figure}
Upon compression, hard sphere glasses undergo a
 {\it Gardner transition} \cite{KPUZ13}; the glass 
basins of configurations split into a 'fractal landscape' of just
marginally stable
metastable states, 
 described 
by  full replica
symmetry breaking (RSB) \cite{MPV87}, 
with soft excitations and divergent
susceptibilities. 
Within this
mean-field scenario, these excitations are responsible for the
anomalous rheological response of amorphous solids \cite{BU16, FS17, JUZY18, JY17, RUYZ15}.  
In the jamming limit this \emph{landscape
marginality} gives rise to the power laws observed in the gaps and
forces distributions and predicts the mechanical properties
of amorphous jammed packings. 

Subsequently, it has been argued \cite{FP16, FPUZ15, FPSUZ17} that
the jamming transition can be thought of as a special case of a satisfiability transition 
 for constraint satisfaction problems \cite{MM11} with continuous
variables (CCSP). In a generic CCSP one seeks
configurations of variables that satisfy a set of constraints. 
From this viewpoint, jamming is the (protocol dependent) point that separates a
satisfiable (SAT) unjammed phase, where the spheres can be arranged to
satisfy the non-overlapping constraints, from an
unsatisfiable (UNSAT) or jammed phase, where some constraints are
violated and spheres overlap.  In this way,
one can generalize the problem of jamming to other
situations.  The simplest one is borrowed from machine learning and is
a non-convex twist of the perceptron classifier \cite{EV01}.  In \cite{FP16} it
has been argued that at the satisfiability transition point (meaning
at jamming) this CCSP displays analogous power law distributions of
gaps and forces whose critical exponents coincide with the ones of
spheres.  Non-trivial generalizations of the perceptron \cite{FHU18, Yo18}
retain the same critical behavior. 

However, the criticality of the gaps and forces distributions 
both in spheres and in the perceptron is generically attributed to the
emergence of jamming and should disappear in the
jammed/UNSAT phase.  This is supported by analytical computations
in the perceptron problem with harmonic cost function and by numerical simulations on
harmonic soft spheres.

In this work we show that this is not always the case: changing the potential or cost function from harmonic to linear, we find jamming criticality in an extensive region of the UNSAT/jammed
phase, far away from jamming.

We consider the UNSAT/jammed phase of the simplest CCSP, the spherical
perceptron model, and we look at local minima of the linear cost function 
(instead of the harmonic one).  
We show that even far from jamming, in an extensive region of the phase diagram, 
the landscape induced by this cost function is non-convex and
composed by metastable minima which are all jamming-critical.
Indeed, for these minima the positive and negative gaps distributions
display power law divergences for small argument: surprisingly, the
critical exponents describing these power laws coincide with the ones
of the jamming transition. Moreover, we find that this behavior is
associated with isostaticity: even when the model is in
the UNSAT phase, there is an extensive number of marginally satisfied
constraints, i.e. constraints that are right at the border of
satisfaction (like perfectly touching spheres).
The jamming-critical phase appears when the number of marginally satisfied constraints equals the number of degrees of freedom of the system which becomes therefore isostatic.

The spherical perceptron with linear cost function has been studied
in \cite{GG91,Gy01,MEZ93} where the phase diagram was obtained using
the replica method and studied at the so-called replica symmetric level in \cite{GG91}. 
While it was known that RSB is needed in the UNSAT/jammed
phase systematic studies beyond 1RSB \cite{MEZ93, GR97, GR00} were not undertaken.
Here we show that the emergence of RSB in the UNSAT phase is related to the emergence
of jamming-like criticality.

{\it The model -- } The spherical perceptron is an optimization
problem defined through an $N$-dimensional vector $\underline w$
on the $N$-dimensional hypersphere $|\underline
w|^2=N$ and by a set of $M=\alpha N$ $N$-dimensional random vectors
$\{\underline \xi^\mu\}$ whose components are \emph{i.i.d.} Gaussian
random variables with zero mean and unit variance. For each of these
vectors one defines the gap $h_\mu$ by
\begin{equation}
h_\mu=\frac 1{\sqrt N} \underline \xi^\mu\cdot \underline w -\sigma\:.
\label{eq_gap}
\end{equation}
We say that a gap $h_\mu$ is (i) satisfied if $h_\mu>0$, (ii) marginally satisfied if $h_\mu=0$ and (iii) unsatisfied if $h_\mu<0$. 
The variables $\sigma$ and $\alpha$ are control parameters.
One can define an energy or cost function associated with the unsatisfied gaps as
\begin{equation}
H[\underline w] = \frac 1p\sum_{\mu=1}^{\alpha N} |h_\mu|^p \theta(-h_\mu)\:
\label{Eq:cost}
\end{equation}
and study its minima.
Eq.~(\ref{Eq:cost}) depends on a parameter $p$ that sets the strength of the cost when gap variables are unsatisfied.
The harmonic perceptron corresponds to $p=2$ and has been studied in \cite{FPUZ15,FPSUZ17}.
Here we set $p=1$ which defines the linear cost function.
This cost function is not very used in soft matter systems
but it is common in the machine learning literature where it is called  
\emph{hinge loss} and plays an important role in Support Vector Machines \cite{ZRT04}.
Furthermore, the case $p=1$ marks the boundary where $H$ passes from a convex ($p>1$) to a non-convex
($p<1$) function of each of the $h_\mu$'s. We stress however that the
convexity of $H$ in the $h_\mu$'s does not necessarily imply convexity of $H$ in the
variables $\underline w$ that live on the hypersphere: 
the loss of convexity is indeed associated with glassiness.

The phase diagram of the spherical perceptron with linear cost function was obtained for $\sigma>0$ in \cite{GG91} (see also \cite{MEZ93})
 and is redrawn in Fig.~\ref{Fig:PD}
in terms of the control parameters $\sigma$ and $\alpha$.
It contains two regions separated by the SAT/UNSAT transition line (red line in Fig.~\ref{Fig:PD}). 
Below this line the problem is SAT (or unjammed)
meaning that with probability one, for $N\to \infty$, 
there are configurations of $\underline w$ for which the
cost function is strictly zero, i.e.
the gaps $h_\mu$ are satisfied for all $\mu=1,\ldots,\alpha N$.
Conversely, above this line the cost function is positive and
an extensive number of gaps are unsatisfied: this is the UNSAT/jammed phase.
In Fig.~\ref{Fig:PD} we plot also the de
Almeida-Thouless (RSB) line \cite{dAT78} (dashed black line):  below this line and in the UNSAT phase, the energy landscape 
is effectively convex and the linear cost function has a unique global minimum. 
Above this line, convexity is lost and multiple metastable minima emerge. 
We are interested in studying the properties of these local
minima.

{\it Numerical Simulations -- }
Local minima of the linear cost function turn out to be non-analytic angular points determined by the intersection of hyperplanes. 
We smooth out the singularity at $h_\m=0$ and define a regularized cost function as
\beq
\begin{split}
H_\e[\underline w]&= \sum_{\mu=1}^{\alpha N} \left|h_\mu+\frac \e 2\right| \theta(-h_\mu-\e) +\frac \z4 (|\underline w|^2-N)^2\\
&+ \frac 1{2\e}\sum_{\mu=1}^{\alpha N} h_\mu^2 \theta(\e+h_\mu)\th(-h_\mu)\\
\end{split}
\label{Eq:cost_smoothed}
\eeq 
where $\e>0$ and   
$\z$ is an arbitrary large constant needed to enforce the spherical constraint $|\underline w|^2=N$.
We implemented the numerical minimization of $H_\e[\underline w]$ using the routine BFGS \cite{BFGS} of the
SciPy library \cite{SciPy}.  For every $\e$, the algorithm
reaches a local minimum. In order to describe the
properties of the linear cost function, we need to study the minima when $\e\to 0^+$.
Therefore, we consider a decreasing sequence of values of $\e$ and minimize the cost function at each step (see SM for details).   
We observe that for $\e$ small enough
there is an extensive fixed set of gaps in the interval
${\cal D} = [-\e, 0]$.  Decreasing $\e$, 
these gaps remain in ${\cal D}$, indicating
that for $\e\to 0^+$ they become marginally satisfied: we
call them contacts in analogy with sphere packings.  
We define the empirical distribution of gaps as
$\rho(h) = \overline{\frac{1}{M} \sum_{\mu=1}^M \delta(h-h_\mu)}$
where the average is taken over many different realizations of the random vectors $\{\underline \xi^\mu\}$.
Therefore, for $\e\to 0^+$ $\rho(h)$ contains a Dirac
delta at $h=0$. Calling ${\cal I}_{\cal D}$ the total number of
contacts, we can define an \emph{isostaticity} index
$c={\cal I}_{\cal D}/N$.  In the inset of Fig.~\ref{Fig:PD} we plot $c$
as a function of $\s$ for $\a=5$.  When replica
symmetry is unbroken (RS),\st{we find} $c<1$ and we say that minima are hypostatic. 
Conversely, when the
minimization is carried in the RSB region we find $c=1$ and therefore we say that 
minima are isostatic.

\begin{figure}
\centering
\includegraphics[width=\columnwidth]{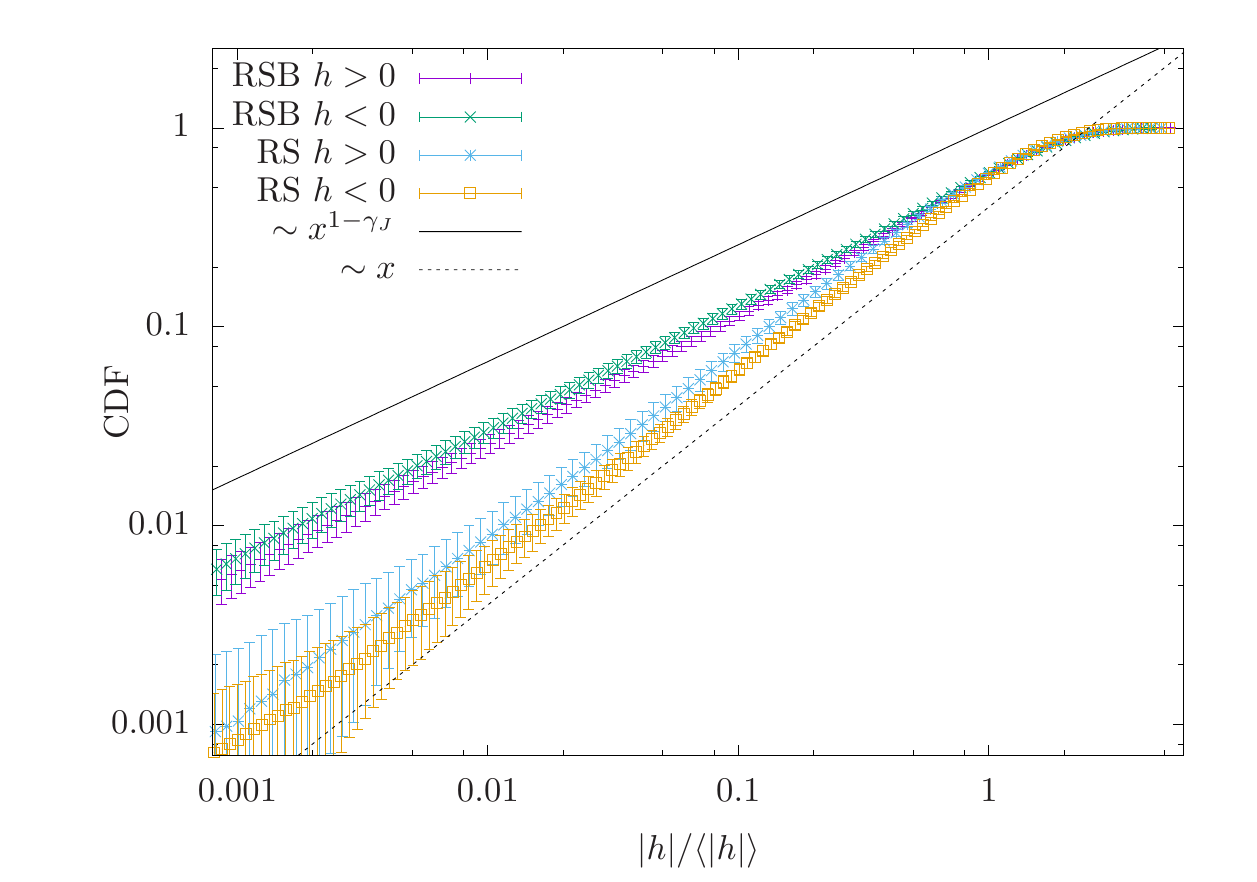}
\caption{The cumulative distribution function (CDF) of both strictly positive and
  strictly negative gaps for $N=1024$ and $\a=5$. We compare the 
distributions in the UNSAT-RSB and UNSAT-RS regions.  The RSB data refer
 to minima at an average energy $H / N =1.01\pm0.02$ and the corresponding value of
$\s$ is $\s=0.219\pm 0.004$. Both positive and negative gaps'
distributions are compatible with a power law at  small argument with
exponent $1-\g_J\simeq 0.59$ (black full line). The RS data refer to minima with $H / N =2.540\pm0.013$ and
$\s=0.757\pm0.010$. One sees there a linear behavior of the CDF,
implying a positive probability density function in the origin.}
\label{Fig:nonzero_gaps}
\end{figure}

\begin{figure}
\centering
\includegraphics[width=\columnwidth]{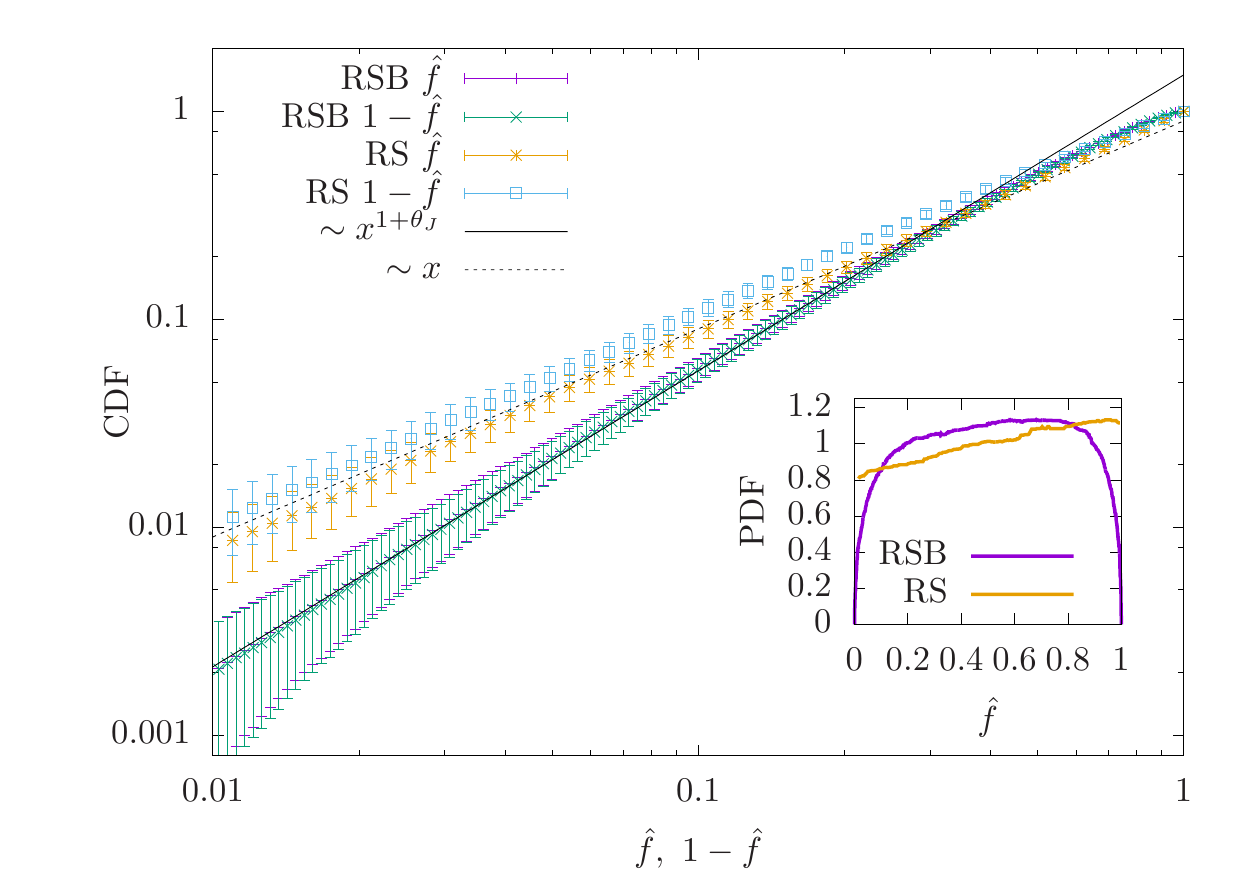}
\caption{The CDF of the virtual forces presented as a function of both $\hat
  f$ and $1-\hat f$ for the same parameters as in
  fig. \ref{Fig:nonzero_gaps}. In the RSB phase the distribution of
  forces vanishes as a power law both in $\hat f=0$ and in $\hat f=1$ (purple line in the inset) and both powers
  are compatible with $\theta_J\simeq 0.42$ corresponding to a CDF
  with a power $1.42$. In the RS region the behavior is linear and the
  PDF is finite at both edges of its support (yellow line in the inset).
}
\label{Fig:zero_gaps}
\end{figure}

Once the contacts are identified we construct the statistics of
strictly positive and negative gaps.  We study what happens to $\rho(h)$ when $h\to 0^\pm$. 
In the RS-UNSAT phase, $\rho(h\to 0^\pm)\to A_{\pm}$
being $A_\pm$ two positive non-universal constants that depend on the control parameters. 
In the RSB region instead
we observe that $\rho(h\to 0^\pm)\sim |h|^{-\gamma_\pm}$
where $\g_{\pm}$ are two critical exponents.
In Fig.~\ref{Fig:nonzero_gaps}, we plot the cumulative
distributions of both positive and negative gaps for minima with an
average energy of $\langle H \rangle/N = 1.01\pm 0.02$, therefore far away from the jamming transition.
Both distributions display 
a non-trivial power law for small
argument.  The critical exponents $\gamma_\pm$ are very
close to each other and have numerical value $\gamma_+\approx \gamma_-\approx 0.41$
which is compatible with the critical exponent $\gamma_J$ 
that controls the positive gaps distribution at jamming transition \cite{CKPUZ14NatComm}.  
Moreover, we can obtain the virtual forces $\hat f$
associated to contacts \cite{FP16}. These are defined
as the Lagrange multipliers needed to enforce that the corresponding
gaps are identically zero \footnote{The virtual forces can be obtained
  as the negative gaps in ${\cal D}$ rescaled by $\e$ for $\e\to 0^+$.}.
  Their empirical distribution 
is defined as
$\rho_f(\hat f) = \overline{\frac{1}{cN} \sum_{i\in {\cal D}} \delta(\hat f-\hat f_i)}$
and has support in $[0,1]$. We find that 
while in the convex phase $\r_f(\hat f)$ is regular at both edges,
in the RSB phase it becomes critical, with pseudogaps close to
both edges of its support, $\r_f(\hat f) \sim \hat f^\th$ and $\r_f(\hat f)\sim (1-\hat f)^{\th '}$ for $\hat f\sim 0^+$ and for $\hat f\sim 1^-$ respectively. 
In Fig.~\ref{Fig:zero_gaps}, we plot the 
cumulative distribution of forces both as a function of $\hat f$ and $1-\hat f$ as obtained from simulations:
we observe
two power laws with exponents $\theta\simeq
\theta'\simeq 0.42$, again compatible with the critical exponent $\theta_J$ that controls the distribution of small forces at jamming.

Therefore numerical simulations show that when the energy
minimization is carried out in the RSB-UNSAT phase, there are three
classes of small gaps: an isostatic set of gaps that are identically zero, and
two sets of positive and negative gaps that accumulate around zero.
Furthermore, the marginally satisfied gaps are associated to a critical distribution of virtual forces.
Unlike for the harmonic case, here
scaling behavior emerges even in the UNSAT phase far from jamming.

{\it Theory -- } 
We analyze the thermodynamic phase diagram of the model using the replica method. Similarly to the case
of the jamming transition in spheres and non-convex perceptron, the
UNSAT critical phase is associated with scaling behavior
that controls the universality of the gaps and forces distributions. 
However, in the case of the jamming transition
there is a single scaling region
that describes small positive gaps and forces. Instead, in the present case 
we also have two additional power laws describing small negative gaps and virtual forces close to one. 
This corresponds to the emergence of an additional scaling region. Both regions can be theoretically
identified from the replica analysis. 
Here we sketch the main steps, details are in the SM. 
The phase diagram and the properties of the model can be
obtained by studying the zero temperature 
limit $\beta=1/T\to \infty$ of its free energy \cite{Ga88}
\begin{equation}
{\rm f}=-\frac 1{\beta N} \overline{ \ln \int {\mathrm d}\underline{w}\, {\mathrm e}^{-\beta H[\underline w]} }
\end{equation}
where the overline stands for the average over the random vectors $\underline \xi^{\mu}$. 
The disorder average can be performed using the replica method \cite{Gy01, FPSUZ17}. 
The RSB phase is described by the probability distribution of the overlap $q=\underline
 w_1\cdot \underline w_2/N$ between different configurations and is captured by the following PDEs \cite{MPV87,
SD84}, valid for real values of $h$ and 
for $q$ in a interval $q\in [q_m, q_M]\subset [0,1]$ determined self-consistently:
{\medmuskip=0mu
\thinmuskip=0mu
\thickmuskip=0mu
\begin{equation}
\begin{split}
\frac{\partial m(q,h)}{\partial q} &= - \frac 12 m''(q,h) - \frac{x(q)}{\lambda(q)} m(q,h)\left[1+m'(q,h)\right]\\
\frac{\partial P(q,h)}{\partial q} &= \frac 12 \left[P''(q,h) -2\frac{x(q)}{\l(q)}\left(P(q,h) m(q,h)\right)'\right]
\end{split}
\label{Eq:ParisiPDE}
\end{equation}
}\noindent where the primes indicate partial derivatives with respect
to $h$ and
the boundary conditions are given by
\begin{equation}
\begin{split}
m(q_{M},h) &= (1-q_M)\frac{\partial}{\partial h} \ln \g_{1-q_M} \star e^{-\b |h|\th(-h)}\\
P(q_m,h) &= \g_{q_m}(h+\s)\:.
\end{split}
\end{equation}
$\g_\D$ is a Gaussian with zero mean and variance $\D$ and $\star $ stands for the convolution operation.
The function $x(q)$ is directly related to the distribution of the overlap $q$ \cite{MPV87} and we have defined
$\l(q) =1-q_M + \int_q^{q_M}\de p\; x(p).$ 
At large $\beta$, one can get the distribution of virtual forces and gaps from the solution of $P(q,h)$ in the limit $q\to 1$.
We analyze Eqs.~(\ref{Eq:ParisiPDE}) in the $\b\to
\infty$ limit in the RSB-UNSAT phase and show that they admit a scaling
solution which accounts for the power laws observed in 
numerical simulations. 
In the UNSAT phase, for $\beta\to \infty $ one
has $q_M\to 1$.
In the limit $0< 1-q\ll 1$ two scaling regimes emerge for $m(q,h)$.
One concerns the region  $h=O( \sqrt{1-q})$, analogue to the one
found at jamming \cite{CKPUZ14NatComm}, and a new one associated to negative gaps for $h= -\hat \lambda(q)+O(
\sqrt{1-q})$, where $\hat \l(q)=\lim_{\beta\to\infty}\b \l(q)\simeq
(1-q)^{(\kappa-1)/\kappa}\gg \sqrt{1-q}$ (being $\kappa<2$ a critical exponent).
Therefore we can write
{
\medmuskip=-1.5mu
\thinmuskip=-1.5mu
\thickmuskip=-1.5mu
\begin{equation}
m(q,h) = \begin{cases}
-\sqrt{1-q} {\cal M}_+\left(\frac{h}{\sqrt{1-q}}\right) &  
|h| 
\sim \sqrt{1-q}\\
-h+\sqrt{1-q} {\cal M}_-\left(\frac{h+\hat
    \lambda(q)}{\sqrt{1-q}}\right) & 
h+\hat \lambda(q)\sim\sqrt{1-q}.
\end{cases}
\end{equation}
}\noindent  
It turns out that the two scaling functions are related by the symmetry relation
${\cal M}_-(t) = t+{\cal M}_+(-t)$.   Moreover, we find that
the scaling function $\MM_+$ satisfies the same equation that 
appear at critical jamming transitions \cite{CKPUZ14JSTAT, FPSUZ17}.
At the same time, the function $P(q,h)$ admits the scaling form
{
\medmuskip=-2mu
\thinmuskip=-2mu
\thickmuskip=-2mu
\begin{equation}
P(q,h)=\begin{cases}
p_+(h) & h\gg \sqrt{1-q}\\
(1-q)^{-a/\kappa}p_0\left(\frac{h}{\sqrt{1-q}}\right) & h\sim \sqrt{1-q}\\
\hat \lambda(q)^{-1} p_-(h\hat \lambda(q)^{-1}) & -h \sim \hat \lambda(q)\\
(1-q)^{- a/\kappa}\tilde p_0\left(\frac{h+\hat \lambda(q)}{\sqrt{1-q}}\right) & |h+\hat \lambda(q)|\sim \sqrt{1-q}\\
\tilde p_+(h+\hat \l(q)) & h+\hat\lambda(q)\ll -\sqrt{1-q}.
\end{cases}
\end{equation}
}\noindent 
The scaling functions $p_+(t)$ and $\tilde p_+(t)$ control
respectively the distribution of small positive and negative gaps.  
Furthermore $\tilde p_0(t)=p_0(-t)$ and $p_0(t)$ satisfies
again a scaling equation that is exactly the same as the one appearing
at critical jamming \cite{CKPUZ14JSTAT, FPSUZ17}. 
This analysis implies that the exponents verify: $\g_{+} = \gamma_- = \gamma_J$ 
and $\theta=\theta'=\theta_J$ being $\gamma_J\simeq 0.41$ and 
$\theta_J \simeq 0.42$ 
the critical exponents controlling the  gaps  and forces distributions at the jamming point of hard spheres.
Finally the nature of the scaling solution implies that the distribution of gaps has an isostatic delta peak of marginally satisfied gaps,
so we get 
 { \medmuskip=0mu \thinmuskip=0mu \thickmuskip=0mu 
\beq 
\begin{split}
\rho(h) &\sim
  \r_+ h^{-\gamma}\th(h) +\r_-(-h)^{-\g}\th(-h) + \frac 1\a \delta(h)\ \ h\to 0
  \end{split}
  \eeq 
  }\noindent 
with $\r_+$ and $\r_-$ two positive constants.

{\it Conclusions -- } We have analyzed the properties of the UNSAT
phase of the spherical perceptron with linear cost function. In the
RS phase the landscape is effectively convex and the global minimum is not critical and hypostatic.  When instead the
minimization is carried out in the RSB phase, we find that local minima are jamming-critical.  They are described by an isostatic number of contacts and the distributions of gaps and virtual
forces follow power laws whose exponents are the same as
the ones characterizing jamming of hard spheres.
We have proposed a 
scaling solution of the RSB equations 
that agrees with the emerging criticality.
There are two clear future directions.  First, it will be interesting to understand what happens to the model if
we change the cost function to a non-convex function of the gaps (i.e. $p<1$ in Eq.(\ref{Eq:cost})).  
Furthermore, it will be interesting to study linear cost
functions in other CCSPs, and see if this leads to universal critical
jammed phases as it happens in the perceptron. We expect that
that this property is generic within mean-field, and our scaling
solution extends to high dimensional spheres, multilayer neural nets
etc. \cite{FHU18,Yo18}. More interesting are problems that are not mean-field in nature. 
For finite dimensional spheres the critical exponents of jamming have been shown to be independent on spatial dimension within
numerical accuracy \cite{CCPZ15}.  
It would be interesting to investigate if the same property holds for the
 jammed phase of linear soft spheres.  We are working in this direction. 
This may provide a finite dimensional physical system
with an extended jamming-critical phase where RSB effects could be
tested.

\paragraph*{ Acknowledgments --} We thank S. Hwang and
J. Rocchi and G. Parisi  for discussions. S.F. and A.S. acknowledge a grant from
the Simons Foundation (No. 454941, Silvio Franz). P.U. acknowledges
the support of \textquotedbl Investissements d'Avenir\textquotedbl  \ LabEx PALM (ANR-10-LABX-0039-PALM) (StatPhysDisSys project).
S.F. is a member of the Institut Universitaire de France (IUF).
This manuscript was partially prepared during P. U.'s visit to KITP and
he acknowledges partial support by the NSF
under Grant No. NSF PHY-1748958.

\appendix
\widetext{
\section{Dictionary between continuous constraint satisfaction problems and jamming of spheres}
In this section we recall the dictionary between jamming of spheres and continuous constraint satisfaction problems with excluded volume constraints.  This mapping has been first presented in \cite{FPUZ15, FPSUZ17}}.\\ 
\begin{itemize}
\item {\bf CCSP.} 
Continuous constraint satisfaction problem with excluded volume constraints can be defined in an abstract way. One considers first a vector $\underline w\in {\rm I\!R}^N$ and a set of $M=\alpha N$ functions $h_\mu(\underline w)\in {\rm I\!R}$ indexed by $\mu=1,\ldots, M$. Each function $h_\mu(\underline w)$ is called \emph{gap}. The constraint satisfaction problem is defined by asking a configuration of $\underline w$ that satisfies all the constraints $h_\mu(\underline w) \geq 0$.
If the problem cannot be satisfied one can define an optimization problem by asking to find a configuration of $\underline w$ that minimizes a cost function. Therefore in general one can define a cost function $H=\frac 1p \sum_\mu |h_\mu|^p\th(-h_\mu)$ where the power $p$ sets the magnitude of the contribution of the negative gaps to the total cost. 
The satisfiable (SAT) phase corresponds to an assignment where $H=0$ while we have $H>0$ in the unsatisfiable (UNSAT) phase. 
The boundary between the two phases is the SAT/UNSAT transition and will be generically referred to as the jamming point. The precise location of this point as a function of the control parameters of the problem may depend on the precise protocol used to obtain solutions. Marginally satisfied gaps are defined by $h_\mu(\underline w)=0$.

\item {\bf Spheres.} In this case the degrees of freedom of the problem are the positions of $N$ spheres in ${\rm I\!R}^d$ denoted by $\{\underline x_i \in {\rm I\!R}^d\}_{i=1,\ldots, N}$. The gap variables are defined for each couple of spheres as $h_{ij} = |\underline x_i - \underline x_j|-\sigma_{ij}$ where $\sigma_{ij} = r_i +r_j$, being $\{r_i\}_{i=1,\ldots,N}$ the values of the radii of the spheres.
The constraint satisfaction problem is defined by asking to find a configuration of the spheres such that $h_{ij}\geq 0$ for all couples $\{ij\}$. A contact between two spheres $i$ and $j$ appears if $h_{ij}=0$ which corresponds to a marginally satisfied gap. 
The jamming point separates the unjammed (SAT) phase where it is possible to find a SAT configuration for the positions of the spheres, from the jammed (UNSAT) phase where the spheres overlap and an extensive number of gaps are negative.
In the jammed phase one can define a cost function $H=\frac1p\sum_{i<j} |h_{ij}|^p\th(-h_{ij})$. For $p=2, 2.5$ one obtain harmonic or Hertzian spheres respectively.

\item {\bf Spherical perceptron.} The degrees of freedom of the spherical perceptron problem are enclosed in an $N$-dimensional vector $\underline w\in {\rm I\!R}^N $ subjected to stay on the sphere $|\underline w|^2=N$. One then introduces a set of $M=\alpha N$ random vectors $\{\underline \xi^\mu\in  {\rm I\!R}^N\}_{\mu=1,\ldots, M}$ whose components are $i.i.d.$ random Gaussian variables with zero mean and unit variance. 
Given these random vectors the gaps are defined by $h_\mu(\underline w) = \underline \xi^\mu\cdot \underline w/\sqrt N -\sigma$ being $\sigma$ a control parameter in the problem playing a role similar to the diameter in spheres. The constraint satisfaction problem requires to find a configuration of $\underline w$ such that $h_\mu\geq 0$ for all $\mu=1,\ldots, M$.
In the UNSAT phase, where such configurations cannot be found, one can define an optimization problem by asking to find configurations of $\underline w$ that minimize a cost function given by $H=\frac 1p \sum_\mu |h_\mu|^p\th(-h_\mu)$. In the present paper we have set $p=1$ that defines the linear cost function but one can also study the harmonic case and this has been done in \cite{FPUZ15, FPSUZ17}. Again, marginally satisfied gaps, analogous to contacts between spheres, correspond to $h_\mu=0$.
\end{itemize}

\section{Scaling solution of the fullRSB equations}
Here we give the details about the scaling solution of the fullRSB
equations we have proposed in the main text to understand the the
critical exponents appearing in the numerical simulations.
Note that in the glassy phase the numerical minimization does not output the ground state of the system.
Therefore we should in principle not expect to have a correspondence between the thermodynamic computation and the numerical findings. Nevertheless we will show that the scaling solution of the RSB equations give rise to critical exponents which are compatible with the numerical findings. 
The same phenomenon happens at jamming and this is another manifestation of universality.
The deep reasons for this is still a matter of active research.

We assume that in the UNSAT RSB region, replica symmetry is broken in a continuous way, at least for $q\to 1$ (see below).
This property implies a proliferation of metastable states in the vicinity of the ground state.
The basic assumption to apply this theory off equilibrium is that this proliferation also happens around local minima reached by local minimization algorithms.

The fullRSB solution of the perceptron problem can be derived following similar steps as in \cite{FPSUZ17} and changing the cost function from harmonic to linear. The free energy of the model can be obtained through a saddle point for an order parameter that is the overlap distribution between different minima in the free energy landscape. 
This quantity is encoded in the function $x(q)$ defined in the
interval $[q_m,q_M]$ where the extrema of the interval, $0<q_m<q_M\leq
1$ \cite{MPV87} have to be determined self-consistently together with
the function $x(q)$.
The saddle point equations for $x(q)$ can be written in terms of two auxiliary functions $m(q,h)$ and $P(q,h)$. In particular one can show that $P(q,h)$ for $q\to q_M$ encodes the properties of the gaps and forces distributions \cite{FPSUZ17}. 
The functions $m(q,h)$ and $P(q,h)$ verify the following PDE's:
\begin{align}
\begin{split}
\dot m(q,h)&= - \frac 12 m''(q,h) - \frac{x(q)}{\lambda(q)} m(q,h)\left[1+m'(q,h)\right]\\
\dot P(q,h)&= \frac 12 \left[P''(q,h) -2\frac{x(q)}{\l(q)}\left(P(q,h) m(q,h)\right)'\right]
\end{split}
\label{Eq:ParisiPDE}
\end{align}
where we have denoted with the dot the derivative w.r.t. $q$ and with
the prime the derivative w.r.t. $h$. 
The boundary conditions of Eqs.~\eqref{Eq:ParisiPDE} are given by
\begin{align}
\begin{split}
m(q_{M},h) &= (1-q_M)\frac{\partial}{\partial h} \ln \g_{1-q_M} \star e^{-\b |h|\theta(-h)}\\
P(q_m,h) &= \g_{q_m}(h+\s)\:.
\end{split}
\end{align}
We want to analyze the behavior of these equations in the zero temperature limit $T\to 0$, $\b\to \infty$.
Therefore we introduce the following re-scaled functions
\begin{align}
y(q) = \beta x(q) \ \ \ \ \ \ \ \ \ \  \hat{\lambda}(q) = \beta(1-q_M) + \int_{q}^{q_M}\de p y(p) \:.
\end{align}
The asymptotic behavior of $m(q_M,h)$ for $q_M\to 1$ is given by
\beq
m(q_M,h) =
\begin{cases}
0,\hspace{1.2cm}  \text{for } h>0\\
-h,\qquad  \text{for } -\b (1-q_M)\ll h\ll 0\\
\b(1-q_M),\quad  \text{for } h\ll-\b(1-q_M)\\
\end{cases}
\label{as_bc}
\eeq
This form suggests to look for an asymptotic solution for the equation for $m(q,h)$ given by
\begin{align}
m(q,h) =
\begin{cases}
0,\hspace{1.2cm}  \text{for } h>0\\
-h,\qquad  \text{for } -\hat{\lambda}(q)\ll h\ll 0\\
\hat{\lambda}(q),\quad  \text{for } h\ll-\hat{\lambda}(q)\\
\end{cases}
\label{limitingM}
\end{align}
Note that Eq.~\eqref{limitingM} agrees with the boundary condition of Eq.~(\ref{as_bc}) and asymptotically satisfies the Parisi equation for $m(q,h)$. 

We now want to study the behavior of the solution of Eqs.~(\ref{Eq:ParisiPDE}) in the limit $T\rightarrow 0$, where $q_M\rightarrow 1$, and then $q\rightarrow 1$.
However we will focus on the scaling limit where $1-q_M \ll 1-q \ll 1$. In this regime, 
based on Eq.~(\ref{limitingM}), we can guess the following scaling form
\begin{equation}
m(q,h) = \begin{cases}
-\sqrt{1-q}\, {\cal M}_+\left(\frac{h}{\sqrt{1-q}}\right) &  |h|\sim \sqrt{1-q}\\
-h+\sqrt{1-q}\, {\cal M}_-\left(\frac{h+\hat \lambda(q)}{\sqrt{1-q}}\right) & h +\hat \lambda(q)\sim \sqrt{1-q}
\end{cases}
\label{scaling_Ms}
\end{equation}
with the boundary conditions:
\begin{align}
\begin{split}
&{\cal M}_+(t\rightarrow\infty) = {\cal M}_-(t\rightarrow\infty) = 0\\
&{\cal M}_+(t\rightarrow-\infty) = {\cal M}_-(t\rightarrow-\infty) = t\:.
\end{split}
\end{align}
These boundary conditions agree with Eq.~(\ref{limitingM}).
In order to compute the scaling equations for $\MM_{\pm}$
we need to consider a scaling form for $y(q)$. We assume that
\begin{align}
\begin{split}
&y(q) \sim y_\chi (1-q)^{-\frac{1}{\kappa}}\\
&1-q_M\sim \chi T^\kappa 
\end{split}
\label{eq:assumption}
\end{align}
so that for $1-q_M\ll 1-q\ll 1$ we get
\begin{align}
\frac{y(q)}{\hat{\lambda}(q)} \sim \frac{\kappa-1}{\kappa} \frac{1}{1-q}\:.
\label{ysul}
\end{align}
Plugging this result inside the Parisi equation for $m(q,h)$ and using the scaling ansatz of Eq.~(\ref{scaling_Ms}) we get the following scaling equations for $\MM_{\pm}$
\beq
\begin{split}
&\begin{cases}
&{\cal M}_+(t) - t {\cal M}_+(t) = {\cal M}_+''(t) + 2 \frac{\kappa-1}{\kappa} {\cal M}_+(t)\left[1-{\cal M}_+'(t)\right]\\
& {\cal M}_+(t\rightarrow\infty) = 0 \qquad  {\cal M}_+(t\rightarrow-\infty)= t\:.
\end{cases}\\
&\begin{cases}
&{\cal M}_-\left(t_-\right) - t_-\ {\cal M}_-\left(t_-\right) = {\cal M}''_-\left(t_-\right) - 2\frac{\kappa-1}{\kappa} \left[t_- - {\cal M}_-\left(t_-\right)\right]{\cal M}'_-\left(t_-\right)\\
&{\cal M}_-(t\rightarrow\infty) = 0 \qquad  {\cal M}_-(t\rightarrow-\infty)= t\:.
\end{cases}
\end{split}
\eeq
The scaling equation for $\MM_+$ coincides with the scaling equation found in the case of jamming of hard spheres and perceptron \cite{CKPUZ14JSTAT, FPSUZ17} and therefore it has the same solution provided that the value of the exponent $\kappa$ coincides with the one appearing at the jamming transition (we will see that this is the case).
Furthermore we can show that the solution of the equation for $\MM_-$ can be obtained from $\MM_+$ using a symmetry transformation.
Indeed it is very easy to show that if $\MM_+(t)$ is a solution of the corresponding scaling equation,
then if we set $\MM_-(t)$ to be
\beq
	{\cal M}_-(t) = t + {\cal M}_+(-t)
	\label{mapping_M}
\eeq
this satisfies its corresponding scaling equation.
Note that the mapping in Eq.~(\ref{mapping_M}) preserves the boundary conditions for $\MM_-(t)$ as it should. 
This tells that the scaling behavior of $\MM_\pm$ can be reduced to just one function ${\cal M}_+(t)$ 
which then happens to be the same as the one controlling the jamming point of spheres \cite{CKPUZ14JSTAT, FPSUZ17}.

We now turn to the analysis of the PDE for $P(q,h)$. 
We consider the following scaling ansatz for $q\rightarrow 1$
\begin{align}
P(q,h)=
	\begin{split}
	\begin{cases}
	p_+(h) & h>0\\
	(1-q)^{-\frac{a}{\kappa}}\ p_0\left(\frac{h}{\sqrt{1-q}}\right) & |h|\sim\sqrt{1-q}\\
	\hat \l(q)^{-1}\ p_-\left(h\hat \l(q)^{-1}\right) & -\hat{\lambda}(q) \ll h \ll -\sqrt{1-q}\\
	(1-q)^{\frac{\tilde{a}}{\kappa}} \tilde{p}_0\left(\frac{h+\hat{\lambda}(q)}{\sqrt{1-q}}\right) & |h+\hat{\lambda}(q)|\sim\sqrt{1-q}\\
	\tilde{p}_+(h)   & h\ll-\hat{\lambda}(q)
 	\end{cases}
	\end{split}
\end{align}
and we have introduced two additional exponents $a$ and $\tilde a$.
The two scaling functions $p_0$ and $\tilde p_0$ live on the same window where the two scaling functions $\MM_\pm$ are different from their
boundary behavior. 
As at the jamming transition \cite{CKPUZ14JSTAT}, we can obtain the corresponding scaling equations by plugging the scaling ansatz into the equation for $P(q,h)$.
In the regime where $|h|\sim\sqrt{1-q}$ we get
\beq
\frac{a}{\kappa}p_0(t) + \frac{1}{2} t\ p_0'(t) = \frac{p_0''(t)}{2} + \frac{\kappa-1}{\kappa} \left[p_0(t)\mathcal{M}_+(t)\right]' \:.
\label{p0_standard}
\eeq	
This equation coincides with the one for the corresponding $p_0(t)$ at the jamming \cite{CKPUZ14JSTAT, FPSUZ17}.
	
Instead, in the regime
where $|h+\hat{\lambda}(q)|\sim\sqrt{1-q}$
we get that	
\beq
\frac{\tilde{a}}{\kappa}\tilde{p}_0(t) + \frac{1}{2}t\ \tilde{p}_0'(t) = \frac{\tilde{p}_0''(t)}{2} - \frac{\kappa-1}{\kappa} \left\{\tilde{p}_0(t)\left[-t+\mathcal{M}_-(t)\right]\right\}' 
\label{scaling_ptilde}
\eeq
The boundary conditions for Eqs.~(\ref{p0_standard}-\ref{scaling_ptilde})
are 
\begin{equation}
\begin{split}
&p_0(t\to \infty)\sim |t|^{-\g}\ \ \ \ \tilde p_0(t\to -\infty)\sim |t|^{-\g'}\\
&p_0(t\to -\infty)\sim |t|^{\theta}\ \ \ \ \tilde p_0(t\to \infty)\sim |t|^{\th'}\\
\end{split}
\end{equation}
which imply the matching conditions
\beq
\begin{split}
p_+(t\to 0^+)\sim t^{-\g} \ \ \ \ \tilde p_+(t\to 0^-)\sim |t|^{-\g'} \\
p_-(t\to 0^-)\sim |t|^{\th} \ \ \ \ p_-(t\to -1^+)\sim |t-1|^{\th'}\:.
\end{split}
\eeq
and the scaling relations
\beq
\begin{split}
\g = \frac{2a}{\k} \ \ \ \ \th=\frac{1-\k+a}{\k/2-1}\\
\g'=\frac{2\tilde a}{\k}\ \ \ \ \th'=\frac{1-\k+\tilde a}{\k/2-1}\:.
\end{split}
\eeq
Using Eq.~\eqref{mapping_M} and the boundary conditions
we can show that the solution for $\tilde p_0(t)$ is given by
\beq
\tilde p_0(t)=p_0(-t)
\eeq
and that $a=\tilde a$.

Up to now, both the scaling functions $\MM_+(t)$ and $p_0(t)$ are not fixed completely since they still depend on the exponent $\k$.
To fix it we follow the same strategy that has been done to construct the scaling solution at jamming. 
We consider the equation
\begin{align}
	\frac{y(q)}{\hat{\lambda}(q)} = \frac{1}{2}\frac{\int dh P(q,h) m''(q,h)^2}{\int dh P(q,h) m'(q,h)^2\left[1+m'(q,h)\right]}
\label{eq_closure}
\end{align}
which can be derived from the fullRSB equations \cite{CKPUZ14JSTAT, FPSUZ17}.
From the scaling forms of $m(q,h)$ and $P(q,h)$ it is easy to see that only the scaling regimes contribute to the integrals.\\
Indeed, at the numerator of the r.h.s. we have
\begin{align}
\begin{split}
	&\int dt p_0(t) \mathcal{M}''_+(t)^2 + \int dt \tilde{p}_0(t) \mathcal{M}''_-(t)^2 =2\int dt p_0(t) \mathcal{M}''_+(t)^2
\end{split}
\end{align}
while at the denominator we have
\begin{align}
\begin{split}
	&\int dt p_0(t) \mathcal{M}_+'(t)^2\left[1-\mathcal{M}_+'(t)\right] + 
	\int dt \tilde{p}_0(t) \left[-1+\mathcal{M}_-'(t)\right]^2\mathcal{M}_-'(t)= 2 \int dt p_0(t) \mathcal{M}_+'(t)^2\left[1-\mathcal{M}_+'(t)\right]
\end{split}
\end{align}
Therefore Eq.\eqref{eq_closure} becomes
\begin{align}
	\frac{\kappa-1}{\kappa} = \frac{1}{2}\frac{\int dt\ p_0(t) \mathcal{M}_+''(t)^2}{\int dt\ p_0(t) \mathcal{M}_+'(t)^2\left[1-\mathcal{M}_+'(t)\right]}
\end{align}
which coincides with the same one that has been found at critical jamming.
Therefore we find that $\k=1.41574$ is the same exponent appearing at the jamming transition
which implies that $\g=\g'=\g_J$ and $\th=\th'=\theta_J$ being $\g_J$ and $\th_J$ the critical exponents 
of the gaps and forces distributions between hard spheres at jamming.

\section{Isostaticity in the RSB-UNSAT phase}

\begin{figure}
\centering
\includegraphics[scale=0.65]{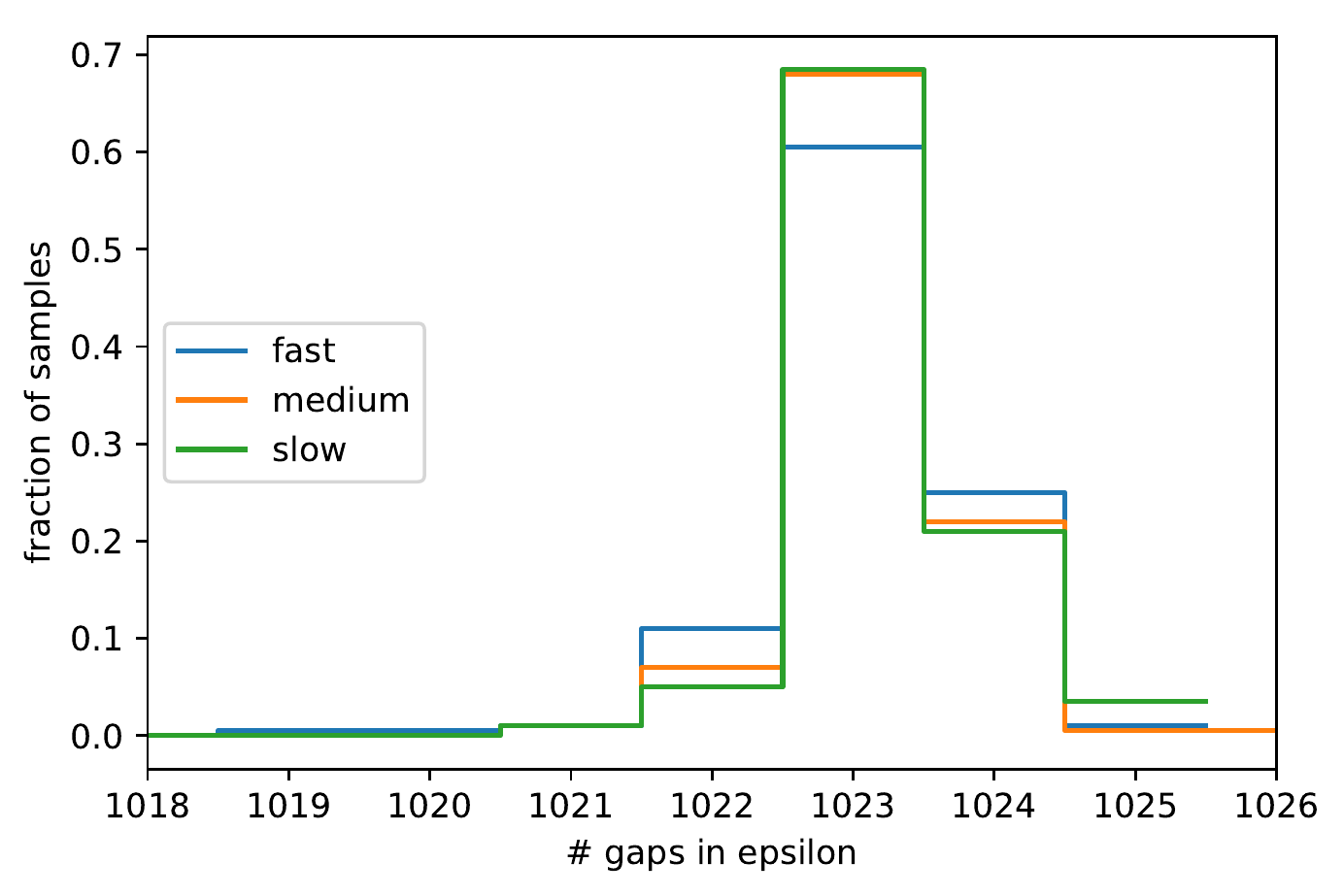}
\caption{The fraction of samples having a given number of gaps in the interval $\DD=[-\varepsilon, 0]$. 
The size of the system is $N=1024$ and the simulations where performed on $200$ samples at $\alpha=5$ and $\sigma=0.22$.
The peak is at the isostatic value $N-1$ and we see deviation from perfect isostaticity of about very few, order one, gaps.
}
\label{Fig:picco_delta}
\end{figure}

In the main text we have underlined that as soon as one enters in the RSB-UNSAT (jammed) phase
of the linear perceptron, the gap distribution contains a delta peak for marginally satisfied gaps.
In order to detect them we consider the smoothed cost function defined in Eq. (3) of the main text and we track the gaps that are contained within the interval $\DD = [-\varepsilon, 0]$ when annealing the smoothing parameter $\varepsilon$.
In Fig.~\ref{Fig:picco_delta} we plot the fraction of samples versus the total number of gaps contained in the interval $\DD$ for $N=1024$ for three different annealing protocols in epsilon:
\begin{itemize}
\item fast: rapid quench where $\varepsilon$ takes two values $\varepsilon=10^{-4}, 10^{-6}$.
\item medium: quench where $\varepsilon$ takes progressively the following values $\varepsilon=10^{-2},10^{-3},10^{-4},10^{-5}, 10^{-6}$
\item slow: $\varepsilon$ takes progressively the following values \\$\varepsilon=10^{-2}, 10^{-3} , 5\cdot 10^{-4}, 10^{-4}, 5\cdot 10^{-5}, 10^{-5}, 7\cdot 10^{-6}, 4\cdot 10^{-6}, 2\cdot 10^{-6}, 10^{-6}$
\end{itemize}
We clearly see a peak at the perfect isostatic value which is $N-1$ (note that the total number of degrees of freedom is $N-1$ because of the spherical constraint on $\underline w$).
Essentially we find that the majority of the samples have exactly $N-1$ gaps in $[-\varepsilon,0]$ which means that they become marginally satisfied for $\varepsilon\to 0^+$. A small number of samples is off from isostaticity for just very few (order one) contacts as it happens at jamming (see for example Sec. II.C of the supplementary information of \cite{CCPZ15}).
Therefore we qualitatively see that fluctuations away from perfect isostaticity are anomalously small as it happens at jamming \cite{HUZ19}.

%

\end{document}